\documentclass[11pt]{article}
\usepackage[tbtags]{amsmath}
\usepackage{amsfonts,amssymb}
\usepackage{graphicx}
\usepackage[all]{xy}
\renewcommand{\baselinestretch}{1.1} % or 2, or whatever

\newcommand{\given}{\mathrel{\vert}\nolinebreak}
\newcommand{\rd}{\textrm{d}}
\newcommand{\bs}[1]{\boldsymbol{#1}}
\newcommand{\COtwo}{\text{CO${}_2$}}
\newcommand{\degC}{\text{${}^\circ$C}}
\newcommand{\Set}[1]{\left\lbrace {#1} \right\rbrace}
\usepackage[modulo]{lineno}
\pdfoutput=1
\usepackage[applemac]{inputenc}
\usepackage[small]{caption}
\title{On the use of simple dynamical systems for climate predictions: \\ 
A Bayesian prediction of the next glacial inception}
\author{Michel Crucifix (1)  and Jonathan Rougier (2) \\ % etc 
\parbox{12cm}{\small{(1) Institut d'Astronomie et de G\'eophysique G. Lema\^\i tre,
Universit\'e catholique de Louvain, 2 chemin du Cyclotron, BE-1348
Louvain-la-Neuve, Belgium (2) Department of Mathematics, University of Bristol, United Kingdom}}}
\date{Published in Eur. Phys. J. Spec. Topics, 174, 11-31 (2009)}
\begin{document}
\bibliographystyle{epj}
\maketitle
\begin{abstract}
%\abstract{\small
Over the last few decades, climate scientists have devoted much effort
to the development of large numerical models of the atmosphere and the
ocean. While there is no question that such models provide important and
useful information on complicated aspects of atmosphere and ocean
dynamics, skillful prediction also requires a phenomenological approach,
particularly for very slow processes, such as glacial-interglacial
cycles. Phenomenological models are often represented as low-order
dynamical systems. These are tractable, and a rich source of insights
about climate dynamics, but they also ignore large bodies of information
on the climate system, and their parameters are generally not
operationally defined. Consequently, if they are to be used to predict
actual climate system behaviour, then we must take very careful account
of the uncertainty introduced by their limitations.  In this paper we
consider the problem of the timing of the next glacial inception, about
which there is on-going debate. Our model is the three-dimensional
stochastic system of Saltzman and Maasch (1991), and our inference takes
place within a Bayesian framework that allows both for the limitations
of the model as a description of the propagation of the climate state
vector, and for parametric uncertainty.  Our inference takes the form of
a data assimilation with unknown static parameters, which we perform
with a variant on a Sequential Monte Carlo technique (`particle
filter').  Provisional results indicate peak glacial conditions in  60,000 years.
\end{abstract}
%} %end of abstract
%
\section{Introduction}
\begin{flushright}
 \linewidth 7cm
\textit{L'analyse math\'ematique peut d\'eduire des ph\'eno\-m\`enes g\'en\'eraux et simples l'expres\-sion des lois de la Nature;
mais l'appli\-cation sp\'eciale de ces lois \`a des effets tr\`es-compos\'es exige une longue suite d'observations exactes.} 
\footnote{ Mathematical analysis allows you to deduce Nature's laws from general and simple phenomena; but applying these laws to highly composite effects requires a long series of exact observations.} \\ Joseph Fourier (1768 -- 1830)
\end{flushright}

This quote by Joseph Fourier appeared first in the `Discours pr\'eliminaire' of the \textit{Analytical theory of heat} \cite{Fourier22aa}. 
%At a time when the reversible Newtonian equations were championed by Pierre-Simon Laplace  (1749 -- 1827) and Joseph Louis Lagrange (1736 -- 1813), the irreversible equations governing heat propagation constituted a genuine mental revolution. 
With this sentence, Joseph Fourier expresses the need for an inductive approach to complex physical phenomena at the macroscopic scale. 
He repeated it at least once, to conclude his \textit{M\'{e}moire sur les temp\'eratures du globe terrestre et des espaces plan\'{e}taires} \cite{Fourier90aa} in which Fourier formulates what is known today as the greenhouse effect. Fourier confesses that ``the question of Earth's temperature is one of the most important and difficult of all the Natural Philosophy''\cite{Fourier90aa}  and solving it was one central motivation for the theory of heat. Clearly, Fourier had fully perceived the complex character of the climate system. How, two centuries later, do we cope with climate's complexity? Which \textit{mathematical analysis} is the most appropriate to get the best out of \textit{observations}?  In this paper we suggest that the most complicated model is not necessarily the most useful. 
Our understanding of the climate system should rather be judged by our ability to build physically consistent models that can be calibrated and validated on observations.

We will illustrate our case with reference to a debate currently taking place in the circle of Quaternary climate scientists. The climate history of the past few million years is characterised by repeated transitions between `cold' (glacial) and `warm' (interglacial) climates. The first modern men were hunting mammoth during the last glacial era. This era culminated around 20,000 years ago \cite{lambeck00} and then declined rapidly. By 9,000 years ago climate was close to the modern one. The current interglacial, called the Holocene, \emph{should} now be coming to an end, when compared to previous interglacials, yet clearly it is not. The debate is about when to expect the next glacial inception, setting aside human activities, which may well have perturbed natural cycles.

On one side, 
Professor Bill Ruddiman carefully inspected and compared palaeo-environmental information about the different interglacial periods. This comparison let him to conclude that glacial inception is largely overdue \cite{ruddiman03,ruddiman07rge}. According to him, the Holocene was not supposed to be this long, but the natural glacial inception process was stopped by an anthropogenic perturbation that began as early as 8,000 years ago (rice plantations and land management by antique civilisations). On the other side, Professor Andr\'{e} Berger and colleagues developed a mathematical model of the climate system, rated today as a `model of intermediate complexity' \cite{gallee91,gallee92} to solve the dynamics of the atmosphere and ice sheets on a spatial grid of $19 \times 5$ elements, with a reasonably extensive treatment of the shortwave and longwave radiative transfers in the atmosphere. Simulations with this model led Berger and Loutre to conclude that glacial inception is not due for another 50,000 years, as long as the \COtwo\ atmospheric concentration stays above 220 ppmv \cite{Berger2002An-exceptionall}.
Who is right? Perhaps both---Crucifix and Berger argued that the two statements are not strictly incompatible \cite{crucifix06eos}.  But perhaps neither.  Both Ruddiman and Berger judge that it is possible to predict climate thousands of years ahead, but is this a realistic expectation? Michael Ghil wondered ``what can we predict beyond one week, for how long and by what methods?'' in a paper entitled \textit{Hilbert problems of the geosciences in the 21st century} \cite{ghil01hilbert}. This is the  motivation for the present article.

\section{Steps towards a dynamical model of palaeoclimates}
\subsection{An inductive approach to complex system modelling \label{remarks}}
A system as complex as climate is organised at different levels : clouds, cloud systems, synoptic waves, planetary waves, pluri-annual oscillations such as El-Ni\~{n}o, glacial-interglacial cycles, and so on.  
% Complex systems and their components act as information processors. This means that their dynamics is such that they can destroy, amplify and even create  information. The difficult mental barrier to overcome for physicists accustomed to Newtonian mechanics is that while the definition of information is subjective (it depends on a \textit{choice} of variables describing the system), the processes of destruction and creation of information rely on general theories.
%The proper of a complex system like climate is that it is impossible to know them fully. It is impossible to know precisely the position and size of any molecule of air and water and to know all the chemical reactions occurring at any given time. 
%In spite of this ignorance, it is possible to make useful predictions about the evolution of a complex system by taking advantage of the existence of organised patterns at a scale much beyond that of the molecule.  
These patterns constitute information that is susceptible of being modelled and predicted.
It is not our purpose to explain here how, in general, patterns emerge in complex systems (\cite{Nicolis07aa} is an up-to-date reference on the subject) but it is useful to have a few notions in mind. 
Schematically, spatio-temporal structures are created by instabilities (necessarily fed by some source of energy), and destroyed by relaxation processes (return to equilibrium).  The balance results in large-scale persistent patterns. % A typical laboratory example would be B\'enard Cells.
In the atmosphere, local hydrodynamical instabilities can result in planetary waves, such as the ones responsible for dominant north-westerly winds in Canada and south-westerly winds in Europe.

In most natural cases, the mechanisms of instability growth are so numerous and intricate that they would completely defeat reductionist modelling approaches for all but the shortest time intervals.  The is certainly true for climate.  
For example, Saltzman repeatedly insisted \cite{saltzman84aa,saltzman02book} on the fact that neither current observations nor modelling of the present state of the atmosphere can possibly inform us of the ice-sheet mass balances with sufficient accuracy to predict their evolution at the timescale of several thousands of years.
But, at the same time, the palaeoclimate archive suggests the behaviour of climate to be fairly regular, over hundreds of thousands of years.  
Consequently, if our interest is in this large-scale behaviour over long time intervals, then we may expect to do better by modelling it directly.  An extreme case of this is simply to fit time-series models.  However, these require detailed and accurate records, and have only a limited ability to account for the influence of time-varying factors, like, in the case of the climate, solar insolation arising from orbital variations.  Between the two extremes of a reductionist model and a time series model there is a huge space for physically-based models that may be empirically tuned.  These are both physical models, in that they embody basic principles in the way that they constrain the interaction between variables, and statistical models, since the coefficients in these models may be treated as uncertain, to be learnt about from observations.  In physics, models that treat observable quantities directly are sometimes termed `phenomenological'.  Our intention is to go further, and embed the phenomenological model in a statistical framework.  This allows us to replace the somewhat haphazard tuning process that these models often undergo with a probability-based updating process, as we will describe in section~3.

\subsection{Empirical evidence about the Quaternary\label{emperical}}
Building a phenomenological model of glacial-interglacial cycles  requires a detailed knowledge of the Quaternary. This section provides a brief overview of the vast amount of information that scientists have accumulated on that period before we tackle Ruddiman's hypothesis. 

\subsubsection{The natural archives}
By the 1920s, geomorphologists were able to interpret correctly the glacial moraines and alluvial terraces as the left-overs of previous glacial inceptions. Penk and Br\"{u}ckner (\cite{Penck09aa}, cited by \cite{berger88}) recognised four previous glacial e\-pochs, named the G\"{u}nz, Mindel, Riss and W\"{u}rm, taking us back about 400,000 years. 
The wealth of data on the Quaternary environments that has since been collected and analysed by field scientists can be appreciated from the impressive four-volume \textit{Encyclopedia of Quaternary Sciences} recently edited by Elias \cite{elias07encyclop}. Analysing and interpreting palaeoenvironmental data involves a huge variety of scientific disciplines, including geochemistry, vulcanology, palaeobiology, nuclear physics, stratigraphy, sedimentology, glacial geology and ice-stream modelling.  Here is a brief summary.

\textbf{Stable isotopes} constitute one important class of natural archives.  
It has been known since the works of Urey \cite{Urey48aa}, Buchanan \cite{Buchanan53aa} and Dansgaard \cite{dansgaard64} that physical and chemical transformations involved in the cycles of water  and carbon fractionate the isotopic composition of these elements. 
For example: ice-sheet water is depleted in oxygen-18 and deuterium compared to sea water; clouds formed at low temperatures are more depleted in oxygen-18 and deuterium than clouds formed at higher temperatures; organic matter is depleted in $^{13}$C, such that inorganic carbon present in biologically active seas and soils is enriched in $^{13}$C ($^{15}$N  is another useful stable palaeo-environmental indicator sensitive to the biological activity of soils).
  The isotopic compositions of  water and biogenic carbon are extracted from deep-sea sediments,  ice and air trapped in ice bubbles, palaeosols, lake-sediments and speleothemes. One of the first continuous deep-sea record of glacial-interglacial cycles was published by Cesare Emiliani \cite{emiliani55}.

  \textbf{Radioactive tracers} are used to estimate the age of the record and rate of ocean water renewal. At the timescale of the Quaternary, useful mother-daughter pairs are $^{230}$Th / $^{238,234}$U (dating carbonates), and $^{40}$K / $^{40}$Ar in potassium-bearing minerals. The ratio 
  $^{230}$Th / $^{231}$Pa is a useful indicator of ocean circulation rates. 

  \textbf{The chemical composition of fossils} is also indicative of past environmental conditions. In the ocean, the amounts of cadmium, lithium, barium and zinc trapped in the calcite shells of foraminifera indicate the amount of nutrients at the time of calcite formation, while the amounts of magnesium and strontium are empirically related to water-temperature.

  Glaciologists have also developed ambitious programmes to analyse \textbf{the composition of air} (oxygen, nitrogen, plus trace gases such as methane, carbon-dioxide and nitrogen oxide, argon and xenon) trapped in ice accumulating on ice sheets, of which the \textit{European Project for Ice Core in Antarctica} (EPICA) is a particularly notable achievement \cite{jouzel07science}.  It was demonstrated that the central plateaus of Antarctica offer a sufficiently stable environment to preserve air's chemical composition over several hundreds of thousands of years. The longest presently available trace gas record --- including carbon dioxide --- goes back to 800\,kyr \cite{Luethi08aa}
%  
%  The chemical composition of water is sensitive to atmospheric circulation patterns and the amount of sea-ice coverage.

  \textbf{Other sources of information.} Plant and animal fossils (including pollens) trapped in lakes, peat-bogs, palaeosols and marine sediments  indicate the then-prevailing palaeoenvironmental conditions. Their presence or absence may be interpreted quantitatively to produce palaeoclimatic maps \cite{climap81}. Preservation indicators of  ocean calcite fossils are used to reconstruct   the history of ocean alkalinity.  Palaeosols and wind-blown sediments (loess) indicate past aridity. The loess grain-size distribution is also sensitive to atmospheric circulation patterns. Geomorphological records are an important source of information about the configuration of past ice sheets, which is complemented by datable evidence (typically coral fossils) on sea-level. 
  \subsubsection{The structure of Quaternary climate changes}

  It is not straightforward to appreciate which aspects of a climate record are relevant to understand climate dynamics at the global scale. For example, minor shifts in oceanic currents may affect the local isotopic composition of water, but with no observable effects on glacial-interglacial cycle dynamics. One strategy is to collect samples from many areas of the world and average them out according to a process called `stacking'. One of the first stacks, still used today, was published by John Imbrie and colleagues \cite{imbrie84} in the framework of the \textit{Mapping spectral variability in global climate project}; it is usually referred to as the \textsc{Specmap} stack. Here we concentrate on the more recent compilation provided by Lisiecki and Raymo \cite{lisiecki05lr04}, called LR04. The stack was obtained by superimposing 57 records of the oxygen-18 composition of benthic foraminifera shells. Benthic foraminifera live in the deep ocean and therefore record the isotopic composition of deep water (an indicator of past ice volume). However, there is an additional fractionation associated with the calcification process, which is related to water temperature. The isotopic composition of calcite oxygen is reported by a value, named $\delta^{18}$O$_c$, giving the relative enrichment of oxygen-18 versus oxygen-16 in calcite (hence the  `c' subscript) compared to an international standard. High $\delta^{18}$O indicates low continental ice volume and/or high water-temperature.

\begin{figure}
\includegraphics{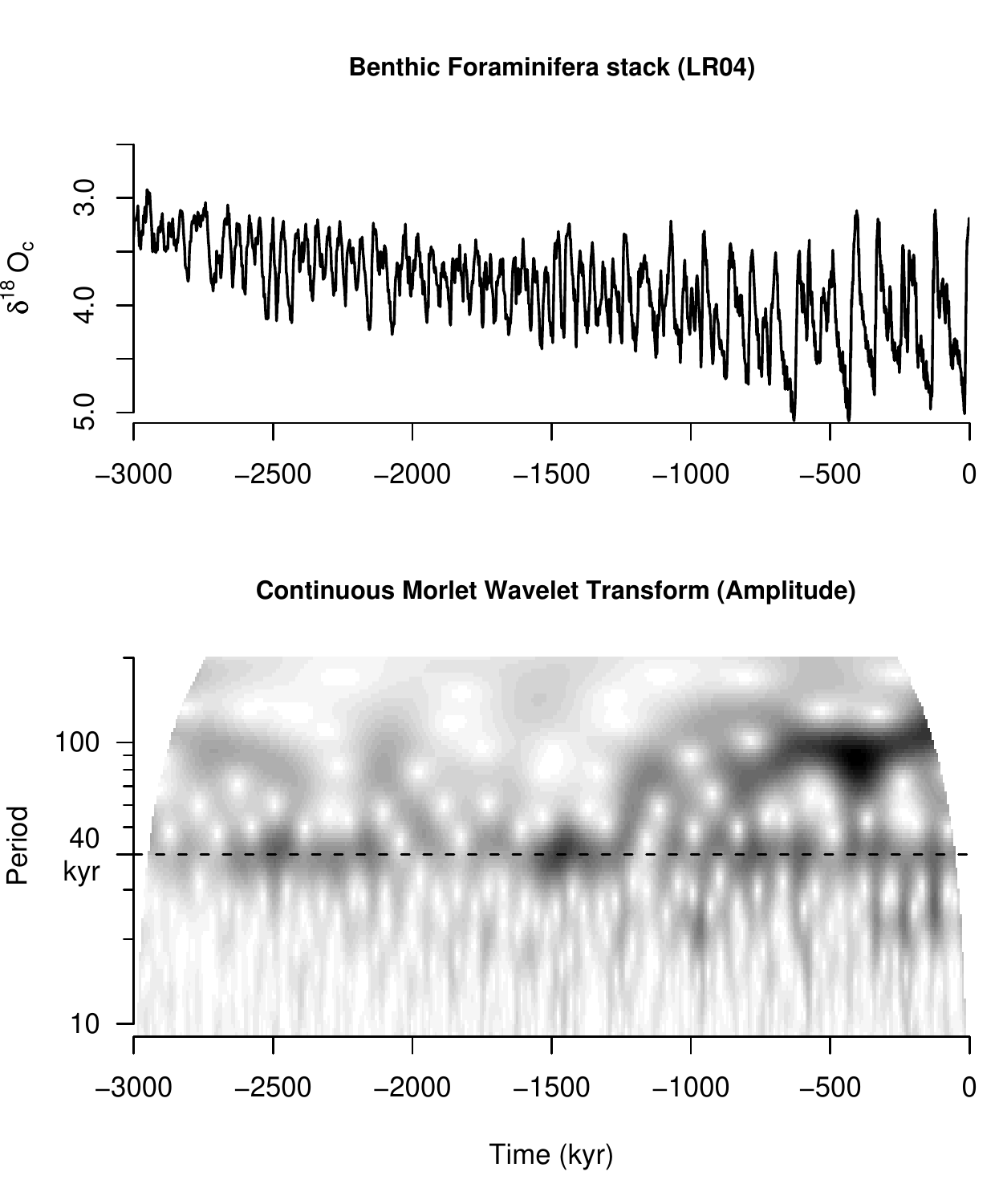}
\caption{
\label{fig:LR04}
The LR04 benthic $\delta^{18}$O stack constructed by the graphic correlation of 57 globally distributed benthic $\delta^{18}$O records \cite{lisiecki05lr04}. Note that the full stack goes back in time to $-5.2$ Myr  (1 Myr = 1 million years, 1kyr=1000 years). The signal is a combination of global ice volume (low $\delta^{18}$O corresponding to low ice volume) and  water temperature (low $\delta^{18}$O corresponding to high temperature). \textit{Top:} The time series, as available from 
\texttt{www.loraine-lisiecki.com}. The vertical axis is reversed as standard practice to get `cold' climates down. 
\textit{Bottom:} Modulus of its continuous Morlet wavelet transform (algorithm of \cite{torrence98}) using $\omega_0=5.4$ but with a normalisation $c(s)=s/\sqrt{\Delta t}$. Large amplitudes are in black.}
\end{figure}

The LR04 stack (Figure \ref{fig:LR04}) shows the gradual transition from the Pliocene---warm and fairly stable---to the spectacular oscillations of the late Pleistocene. The globally averaged temperature at the early Pliocene was about 5\degC\ higher than today (\cite{raymo96aa} and references therein); that at the last glacial maximum (20,000 years ago) was about 5\degC\ lower. Key research questions are to characterise these oscillations, understand their origin and quantify their predictability. 

The Morlet Continuous wavelet transform \cite{grossman84} provides a simple summary of the backbone of these oscillations. The LR04 record is dominated for most of the time by a 40,000~yr signal until roughly 900,000 years ago, after which the 40,000~yr signal is still present but topped by longer cycles. At the very least, this picture demonstrates that LR04 contains structured information susceptible of being modelled and possibly predicted.

How complex a model is needed to model such a series?  There is no clear-cut answer. Time-series extracted from complex systems are sometimes characterised by their \textit{correlation dimension}, which is an estimator for the fractal dimension of the corresponding attractor \cite{GRASSBERGER83aa}. The first estimates for the Pleistocene were provided by Nicolis and Nicolis \cite{nicolis86} ($d=3.4$) and Maasch et al.\ \cite{Maasch89aa} ($4\le d \le 6$). For this article we calculated correlation dimension estimates for the LR04 stack ($d \approx 2.4$) and the HW04 stack \cite{huybers04Pleistocene} ($d \approx 2.5$); estimates were made using the `fractal' R package \cite{Constantine07aa}. HW04 is  similar to LR04 but it is based on different records and dating assumptions. Note, however, that several authors, including Grassberger \cite{GRASSBERGER86aa,Pestiaux84aa,VAUTARD89aa}, have discouraged the use of correlation dimension estimates for the `noisy and short' time series typical of the Quaternary because they are  overly sensitive to sampling and record length. %They are therefore unreliable.%and they concluded that it is not possible to reliably distinguish stochastic from low-order chaotic noise in such series.

\begin{figure}[t]
  \begin{minipage}{0.55\columnwidth}
\includegraphics{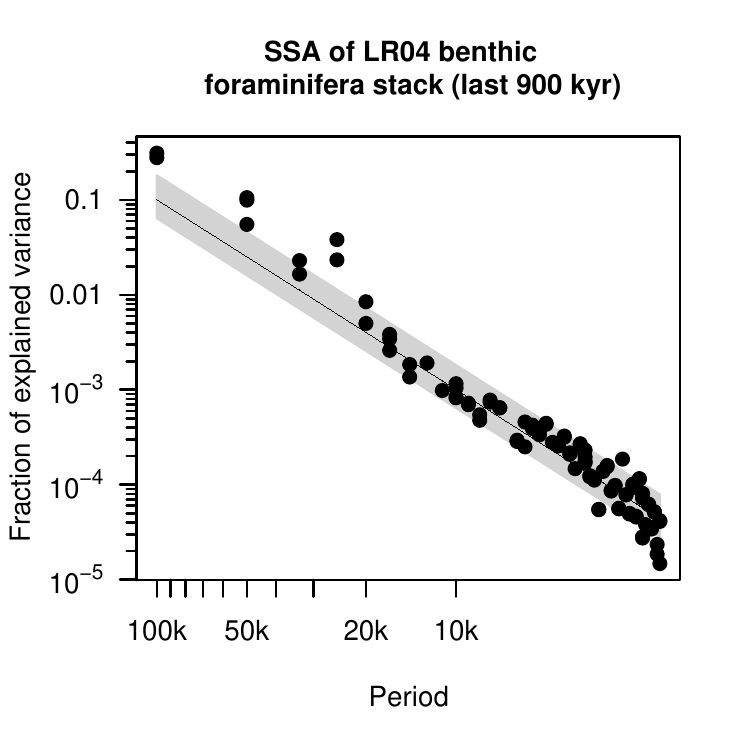}
  \end{minipage}\hfill
  \begin{minipage}{0.40\columnwidth}
\caption{
\label{fig:LR04ssa}
Singular Spectrum Analysis  (SSA) of the LR04 benthic stack. The figure shows the eigenvalues of the lagged-covariance matrix of rank $M=100$ as given by \cite{Ghil02aa}, eq.~(6).   The record was cubic-spline interpolated ($\Delta t = 1$kyr) and only the most recent 900~kyr were kept. Eigenvalues are here plotted according to their characteristic period. They then appear to be broadly consistent with a  1st-order autoregressive process (the AR1-null hypothesis is represented by the grey shade \cite{Shun99aa,Gudmundsson08aa}), but significant features do appear at periods of 20~kyr and longer. Note that the spectrum is conditioned by astronomical tuning assumptions, as may be observed by comparison with non-tuned records.}
\end{minipage}
\end{figure}

In response to this problem Ghil and colleagues  \cite{VAUTARD89aa,YIOU94aa}
advocated singular-spectrum analysis (SSA), in which a time series is linearly decomposed into a number of prominent modes (which need not be harmonic), plus a number of small-amplitude modes. Assuming  that the two groups are separated by an amplitude gap, the first group provides the low-order backbone of the signal dynamics while the second group is interpreted as noise. SSA was applied with some success to various sediment and ice-core records of the last few glacial-interglacial cycles \cite{YIOU94aa} and has in general confirmed that the backbone of climate oscillations may be captured as a linear combination of a small number of amplitude and/or frequency-modulated oscillations, plus a long-term trend. SSA of the last 900\,kyr of LR04 (Figure \ref{fig:LR04ssa}) confirms this statement, but ordering eigenvalues according to their characteristic period indicates that the `noisy background' of LR04 is essentially organised as first-order autoregressive process.

\subsubsection{The Achilles heel of reconstructions}
Now we must mention a particularly difficult and intricate issue: dating uncertainty in palaeoclimate records. No palaeoclimate record is dated with absolute confidence. Marine sediments are coarsely dated by identification of a number of reversals of Earth's magnetic field, which have been previously dated in rocks by radiometric means (\cite{Raymo97aa} and references therein). Magnetic reversals are rare (four of them over the last 3 million years) and their age is known with a precision no better than 5,000 years. Local sedimentation rates may vary considerably between these time markers, such that any individual event observed in any core taken in isolation is hard to date.  Irregularities in the sedimentation rate blur information in the spectral analysis. 

One strategy to tackle this issue is to assume synchrony between oscillation patterns identified in different cores. Statistical tests may then be developed on the basis that dating errors of the different cores are independent. For example, Huybers (2007) \cite{huybers07obl} considered the null-hypothesis that glacial-to-interglacial transitions (they are called \textit{terminations} in the jargon of palaeoclimatologists) are independent of the phase of Earth's obliquity. While this null-hypothesis could not  be rejected on the basis of a single record, the combination of 14 cores allowed him to reject it with 99\% confidence, proving once more the effect of the astronomical forcing on climate (see section~\ref{sec:milan}).  %First tests of this kind were carried out by Hays et al. in a seminal paper \cite{hays76}.
Note statistical tests to assess the significance of a correlation between two ill-dated palaeoclimate records are only now being developed \cite{Haam09aa}.

\begin{figure}[t]
\includegraphics[width=\textwidth]{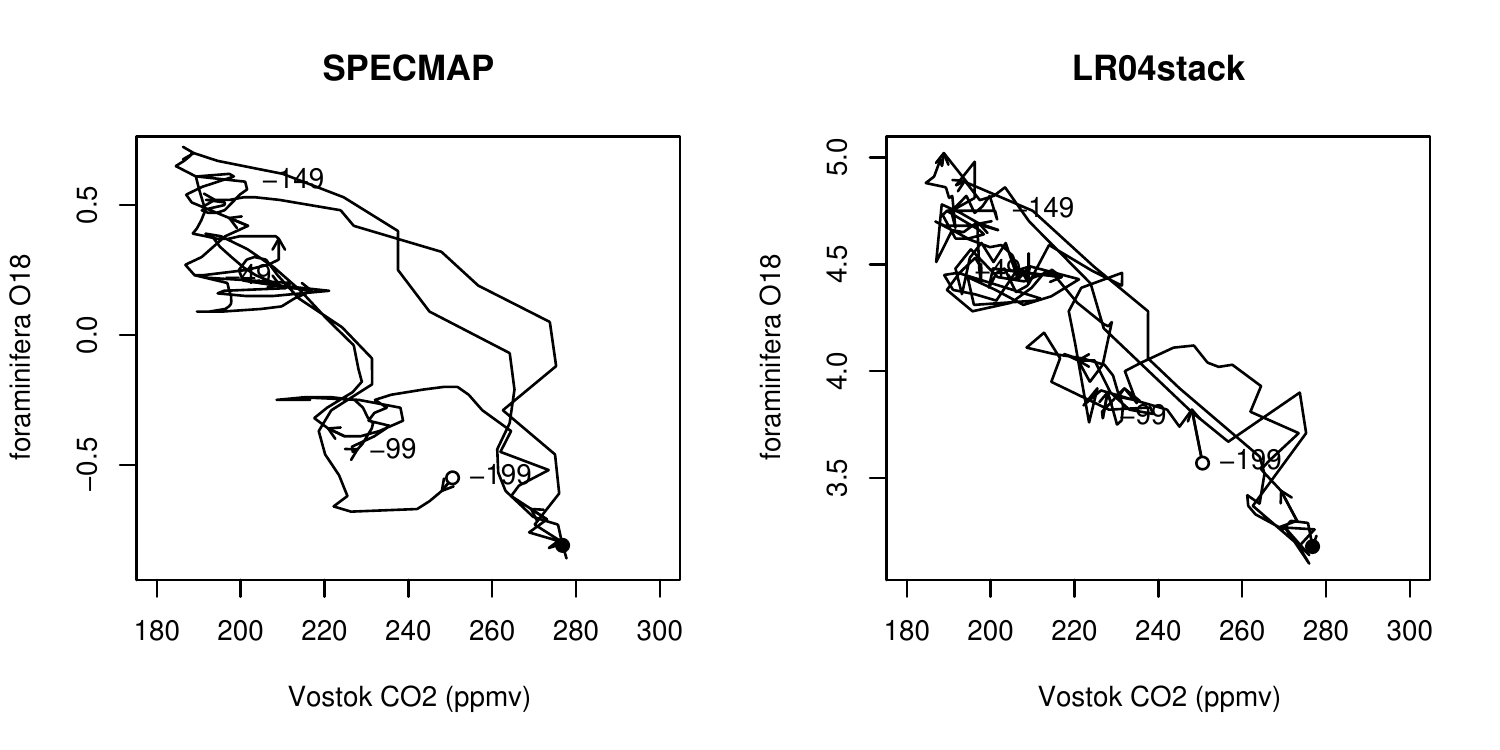}
\caption{The concentration in \COtwo\ measured in the Vostok ice core record \cite{petit99} over the last 200~kyr is plotted versus two proxies of continental ice volume.  Left: the planctonic $\delta^{18}$O stack by Imbrie et al.~(1984) (standard units). Right: the benthic $\delta^{18}$O stack by Lisiecki and Raymo~(2004) (per mil). Numbers are dates, expressed in kyr BP (before present).  The Imbrie stack suggests hysteresis with \COtwo\ leading ice-volume, but  LR04 does not show this so clearly.}
\label{fig:phase}
\end{figure}

Another strategy is known as \textit{orbital tuning}. 
The method consists in 
squeezing  or stretching the time-axis of the record to match the evolution of one or a combination of orbital elements, possibly pre-filtered by a climate model  \cite{imbrie84,hays76}. This strategy has undeniably engendered important and useful results (e.g. \cite{SHACKLETON90aa}), but at the cost of imposing link between orbital forcing and the record.  Experienced investigators recognise that orbital tuning has somehow contaminated most of the dated palaeoclimate records available in public databases. % This has increased the risk of tautological reasoning about the nature of glacial cycles.
%
%For example, compare the two SSA analyses shown in Figure~\ref{fig:LR04ssa}. As we mentioned above, LR04 and HW04 are two stacks of the Pleistocene but LR04 is made of more records (57 instead of 21 in HW04) and it is astronomically tuned.  We can see from the SSA analysis that LR04 presents more quasi-periodic structures than HW04 (quasi-periodic modes are identified as pairs of eigenvalues with approximately the same amplitude). Why is this the case? Is it because age errors in HW04 blurred the interesting information, or is it because this information has been artificially injected into LR04 by the tuning process? There is probably a bit of both (but note that HW04 displays a similar wavelet structure to LR04).
%
Quantification of leads and lags between \COtwo\ and ice volume in particular is hostage to hidden dating assumptions and circular reasoning.
Here is one typical illustration.  Saltzman and Verbitsky
 showed on several occasions (e.g.\ \cite{SALTZMAN94ab})
 a phase diagram showing the SPECMAP  $\delta^{18}$O stack versus the first full ice-core records of \COtwo\ from Vostok \cite{barnola87,jouzel93}. It is reproduced here (Figure \ref{fig:phase}, left). The phase diagram clearly suggests that \COtwo\ leads ice volume at the 100~kyr time scale. However, a detailed inspection of the original publications reveals that the SPECMAP record was astronomically tuned, and that the Vostok time-scale, called GT4, uses a conventional date of isotopic stage 5.4 of 110~kyr BP \ldots\ by reference to SPECMAP! \cite{jouzel93}.  The observed hysteresis may therefore be an artifact. The situation today is that there is no clear consensus about the phase relationship between ice volume and \COtwo\ at the glacial-interglacial time scale (compare \cite{Kawamura07aa,Ruddiman03aa,shackleton00}). According to Ruddiman's analysis \cite{Ruddiman03aa}, \COtwo\ leads ice volume at the precession (20~kyr) period, but \COtwo\ and ice volume are  roughly synchronous at the obliquity (40~kyr) period.  Current evidence about the latest termination is that the decrease in ice volume and the rise in \COtwo\ began around 19,000 years ago \cite{Kawamura07aa,yokoyama00lgm}.

\subsection{Getting physical laws into the model}\label{sec:milan}
So far we have learned that palaeoclimate oscillations are structured and that it is not unreasonable to attempt to model them with a phenomenological model forced by the astronomical variations of Earth's orbit. 
What is the nature of the physical principles to be embedded in such a phenomenological model, and how can they be formalised? The history of Quaternary modelling is particularly enlightening in this respect (\cite{berger88} provides an extensively documented review of Quaternary climates modelling up to the mid-1980s).

Joseph Adh\'{e}mar (1797--1862) \cite{adhemar1842} first suggested that the cause of glaciations is the precession of the equinoxes, but subsequently there was disagreement about how precession affected climate.  Joseph John Murphy maintained that cold summers (occurring when summer is at aphelion) favour glaciation \cite{Murphy76aa}, while James Croll (1821--1890) maintained that cold winters are critical \cite{croll1875}.
Croll's book demonstrates a phenomenal encyclopaedic knowledge.  His judgements were in places particularly far-sighted, but they are barely substantiated by the \textit{mathematical analysis} Fourier was so insistent about.  
The nature of his arguments was essentially empirical, if not in places frankly rhetorical. 

Milutin Milankovitch (1879--1958) is today acknowledged as the pioneer of mathematical climatology. In a highly technical book that crowned a series of articles written between 1920 and 1941 \cite{milankovitch41}, Milankovitch extended Fourier's work to estimate the zonal distribution of Earth's temperature from incoming solar radiation. He also computed the effects of changes in precession, eccentricity and obliquity on incoming solar radiation at different latitudes to conclude, based on geological evidence, that summer insolation drives glacial-interglacial cycles, thus supporting Murphy.

Mathematical analysis allowed Milankovitch to deduce the consequences of certain fundamental principles, such as the laws of Beer, Kirchhoff and Stefan, on global quantities such as the Earth's temperature. But Milankovitch also used empirical macroscopic information, such as the present-day distribution of the snow-line altitude versus latitude, to estimate the effects of temperature changes on the snow cover. In today's language, one may say that Milankovitch had accepted that some information could not be immediately inferred from microscopic principles because it depended on the way the system as a whole responded to its numerous and intricate constraints (Earth's rotation, topography, air composition etc.).

The marine-record study published by Hays, Imbrie and Shackleton \cite{hays76} is often cited as the most indisputable proof of Milankovitch's theory. Hays et al.\  identified three peaks in the spectral estimate of climate variations that precisely correspond to the periods of obliquity (40~kyr) and precession (23~kyr and 19~kyr) calculated analytically by Andr\'{e} Berger\footnote{The supporting papers by Berger would only appear in the two following years \cite{berger77,Berger77ber2,berger78}. Hays et al.\ based their analysis on a numerical spectrum estimate of the orbital time-series provided by Vernekar \cite{vernekar72}.}. 
 Milankovitch's theory of ice ages, though, had to be revisited in the face of evidence---already available in an article by Broecker and van Donck \cite{broecker70}---that the glacial cycle is 100,000 years long, ice build-up taking about 80,000 years and termination about 20,000 years \cite{hays76,broecker70}. Neither the 100,000-year duration of ice ages, nor their saw-tooth shape were predicted by Milankovitch.
The bit Milankovitch's theory is missing is the \textit{dynamical aspect} of climate's response. Glaciologist Johannes Weertman \cite{weertman76} consequently addressed the evolution of ice sheet size and volume by means of an ordinary differential equation, thereby opening the door to the use of dynamical system theory for understanding Quaternary oscillations.
%The works led by Ghil (e.g.: \cite{ghil81}) and Saltzman \cite{saltzman02book} constitute important references in this respect.

In the meantime, general circulation models (GCMs) of the atmosphere and oceans  running on supercomputers became widely available (cf.\ \cite{randall00} for a review), and 
used for palaeoclimate purposes \cite{broccoli87lgm,kutzbach81,mitchell93}. %These models provide a consistent picture of the planetary dynamics of the atmosphere and the oceans. 
Just as Milankovitch applied Beer and Kirchoff's laws to infer Earth's temperature distribution, GCMs allow us to deduce certain aspects of the global circulation from our knowledge of balance equations (i.e.\ representing continuity and conservation) in each grid cell. 
However, these balance equations are highly uncertain at the resolution of current solvers--- clouds, for example, are very hard to represent reliably---and quantifying this uncertainty at the global scale is a difficult problem which is only now being systematically addressed \cite{allen00uncertainty,murphy04qump}.
While GCMs are undeniably useful to constrain the immediate atmospheric response to changes in orbital parameters, they are far too uncertain to estimate glacial accumulation rates reliably enough to predict the evolution of ice sheets over tens of thousands of years \cite{saltzman84aa}. 

In the following sections we will concentrate on a three-dimensional dynamical climate  model proposed by Barry Saltzman. This choice was guided by the ease of implementation as well as the impressive amount of supporting documentation \cite{saltzman02book}. 
However, there were alternatives to this choice. The reader is referred to the article by Imbrie et al.\ \cite{Imbrie92aa} and pp.\ 264--265 of Saltzman's book \cite{saltzman02book} for a summary with numerous references organised around the dynamical concepts proposed to explain glacial-interglacial cycles (linear models, with or without self-sustained oscillations, stochastic resonance,  model with large numbers of degrees of freedom).

Among the alternatives to Saltzman's model, the series of models published by Ghil and colleagues \cite{kallen1979,ghil81,letreut83} have some of the richest dynamics. They present self-sustained oscillations with a relatively short period (6,000 years). The effects of the orbital forcing are taken into account  by means of a multiplicative coefficient in the ice mass balance equation. This causes non-linear resonance between the model dynamics and the orbital forcing.  The resulting spectral response presents a rich background with  multiple harmonics and band-limited chaos. More recently, Gildor and Tziperman \cite{gildor01ab} proposed a model where sea-ice cover plays a central role. In this model, termination occurs when extensive sea-ice cover reduces ice accumulation over ice sheets. Like Saltzman's, this model presents 100~kyr self-sustained oscillations that can be phase-locked to the orbital forcing.
We believe that much is to be learned from a systematic analysis of these simple models, and they easily find their place in the climate modelling hierarchy proposed by Held \cite{held05hierarchy}.

Scientists with long field experience have also proposed models.  These are usually qualified as `conceptual', in the sense that  they are formulated as a worded causal chain inferred from a detailed inspection of palaeoclimate data without the support of differential equations. Good examples are \cite{Ruddiman03aa,Imbrie92aa,imbrie93,Ruddiman06aa}. In the two latter references, Ruddiman proposes a direct effect of precession on \COtwo\ concentration and tropical and southern-hemisphere sea-surface temperatures, while obliquity mainly affects the  hydrological cycle and the mass-balance of northern ice sheets. 

\section{The Saltzman model (SM91)}
\subsection{Outline}
As a student of Edward Lorenz, Barry Saltzman (died 2002)  contributed to the formulation and study of the famous Lorenz63 dynamical system \cite{lorenz63} traditionally  quoted as the archetypal low-order chaotic system\footnote{The \textit{Acknowledgments} of the Lorenz (1963) paper read: ``The writer is indebted to Dr. Barry Saltzman for bringing to his attention the existence of nonperiodic solutions of the convection equations.''}. Saltzman was therefore in an excellent position to appreciate the explanatory power of dynamical system theory. Between 1982  and 2002 he and his students published several dynamical systems deemed to capture and explain the dynamics of Quaternary oscillations \cite{saltzman84aa,saltzman02book,Saltzman82aa,saltzman90sm,Saltzman91sm,saltzman93}. In the present  article we choose to analyse the `palaeoclimate dynamical model' published by Saltzman and Maasch (1991) \cite{Saltzman91sm}.  We will refer to this model as SM91.

Saltzman judged that the essence of Quaternary dynamics should be captured by a three-degree-of-freedom dynamical system, possibly forced by the variations in insolation caused by changes in orbit \cite{saltzman84aa}. The evolution of climate at these time scales is therefore represented by a trajectory in a 3-dimensional manifold, which Saltzman called the ``central manifold''. The three variables are ice volume ($I$), atmospheric CO$_2$ concentration ($\mu$) and deep-ocean temperature ($\theta$). It is important to realise that Saltzman did not ignore the existence of climate dynamics at shorter and longer time scales than those that characterise the central manifold, but he asserted the hypothesis that these modes of variability may be represented by \textit{distinct} dynamical systems. 
In this approach, the fast relaxing modes of the complex climate system are in statistical equilibrium with its slow and unstable dynamical modes. This is sometimes called the `adiabatic elimination' \cite{Haken81aa}.

A statistical representation of the fast modes is plainly justified when a spectral gap separates slow and fast variances. Unfortunately ---  and contrary to a common perception \cite{Kutzbach1976471,Shackleton90ab}  --- there is no evidence for such a gap in the palaeoclimate record \cite{Ditlevsen96aa,Pelletier1998157,Huybers06aa}. However, the latter references 
document 
a change in the power-law exponent of the spectral background: signal energy decays faster with frequency below $(100 \textrm{yr})^{-1}$ than above.
This is evidence that  the effective dissipation time-scale is larger for high frequencies, and that the dynamics of slow and fast climatic oscillations are at least  partly decoupled. 
To be complete, note that \cite{Ditlevsen96aa} contrasted atmospheric variability during modern glacial era on the basis Greenland ice core data, and found that atmospheric variance was larger, and fatter-tailed during the glacial era.
\subsection{Formulation}

% \subsection{The Saltzman model (SM91): Equations}
The ice mass-balance is the result of the contribution of four terms: a drift, a term inversely proportional to the deviation of the summer mean temperature at high latitudes ($\tau$) and  a relaxation term 
%, and a stochastic forcing representing ``\textit{all aperiodic phenomena not adequately parameterised by the first three terms}'':
%
\begin{equation}
\frac{dI}{dt}=\varphi_1 - \varphi_2 \tau - \varphi_3 I.
\label{dIdt}
\end{equation}
By diabatic elimination, $\tau$ is in thermal equilibrium with the slow variables $[I,\mu,\theta]$ and its mean may therefore be estimated as a diagnostic function of the latter. Saltzman adopted a linear approximation of the form:
\begin{equation} 
\tau= \tau_0 + \alpha_I (I-I_0) + \alpha_\mu(\mu-\mu_0) + \alpha_\theta (\theta-\theta_0) + \alpha_R R,
\label{tau}
\end{equation}
where $R$ is the astronomical forcing. Saltzman used incoming insolation at 65$^o$ N at summer solstice, which can be calculated using the Berger algorithm \cite{berger78}.  The reference state $[I_0,\mu_0,\theta_0]$ will be defined later on.

The CO$_2$ equation includes the effects of ocean outgassing as temperature increases, a forcing term representing the net balance of CO$_2$ incorporated into the atmosphere minus that eliminated by silicate weathering, and a non-linear dissipative term:
\begin{eqnarray}
\frac{d\mu}{dt}&=&\beta_0 - \beta_{\theta} \theta + F_\mu - K_\mu \mu \quad \label{dmudt} 
\textrm{with} \quad K_\mu  =  \beta_1 - \beta_2 \mu + \beta_3 \mu^2. 
\end{eqnarray}
The dissipative term ($K_\mu \mu$) is a so-called Landau form and its injection into the CO$_2$ equation is intended to cause instability in the system. In earlier papers (e.g. \cite{saltzman88}), Saltzman [and Maasch] attempted to justify similar forms for the CO$_2$ equation on a reductionist basis: each term of the equation was identified  to specific, quantifiable mechanisms like the
effect of sea-ice cover on the exchanges of CO$_2$ between the ocean and the atmosphere, or the ocean circulation on nutrient pumping.  It is noteworthy that Saltzman gradually dropped and added terms to this equation (compare \cite{saltzman90sm,Saltzman91sm,saltzman88})  to arrive at the above formulation in which they posit a carbon cycle instability without enquiring too deeply into causal mechanisms.

The deep-ocean temperature simply assumes a negative dependency on ice volume with a dissipative relaxation term:
\begin{equation}
\frac{d\theta}{dt}=\gamma_1 - \gamma_2 I -\gamma_3\theta. 
\label{dthetadt}
\end{equation}

Without loss of generality we are now free to assume that $[I_0,\mu_0,\theta_0]$ is a fixed point of the evolution equations. Then evolution equations for $[I',\mu',\theta'] \equiv [I,\mu,\theta] - [I_0,\mu_0,\theta_0]$ may be expressed as follows:
\begin{subequations}
\begin{align}
\frac{dI'}{dt}&=-a_1 [ k_\mu \mu' + k_\theta \theta' + k_R R'(t)] - K_I I' \\
\frac{d\mu'}{dt}&=b_1\mu'-b_2\mu'^2-b_3\mu'^3-b_\theta \theta'  \\
\frac{d\theta'}{dt}&=-c_1I' - K_\theta' \theta' 
\end{align}
\label{SM91}
\end{subequations}
%This leaves us with us with six parameters for the system determining the trajectory of  $\{I',\mu',\theta'\}$. 
The different parameters $a_1,b_i,c_1,k_i$ are functions of the $\alpha_i,\beta_i,\gamma_i,I_0,\mu_0$ and $\theta_0$. Some of these we will treat as known and fixed, but others we will treat as unknown, as available to be tuned to observations. For this we need a formal statistical framework that unifies the SM91 model and the climate system that the model purposes to represent.

\section{The Bayesian inference process \label{bayes}}

\begin{figure}
 % \begin{minipage}{0.55\columnwidth}
\includegraphics{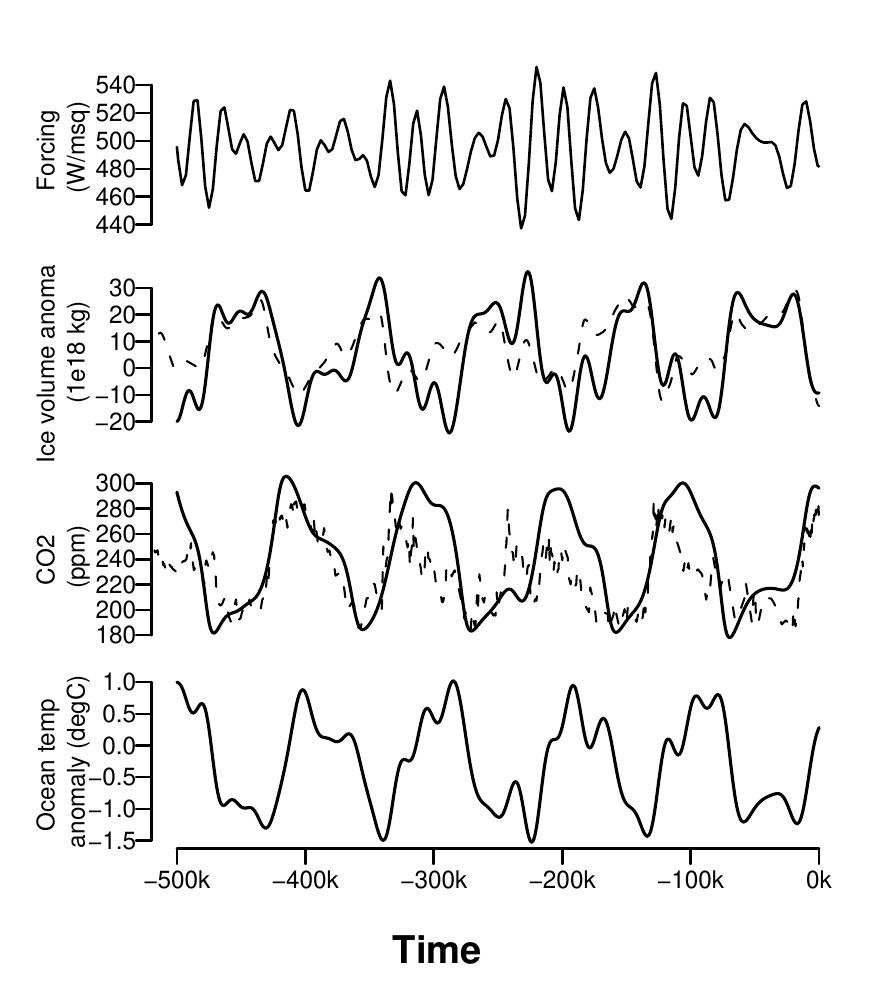}
%  \end{minipage}\hfill
%  \begin{minipage}{0.40\columnwidth}
\caption{Response of the palaeoclimate model of Saltzman and Maasch (1991) \cite{Saltzman91sm}. Shown are the insolation forcing, taken as the summer solstice incoming solar radiation at 65$^\circ$ N \cite{berger78}; the ice volume anomaly (full), overlain with the SPECMAP planctonic $\delta^{18}O_c$ stack \cite{imbrie84} (dashed), the CO$_2$ atmospheric concentration, overlain with the Antarctic ice core data from Vostok and EPICA \cite{Luethi08aa,petit99}, and finally deep-ocean temperature. 
SPECMAP foraminifera data are scaled linearly assuming that the transition between the last glacial maximum and today represents a melt-down of $44\cdot10^{18}$~kg of ice \cite{lambeck00,marsiat90} (this scaling is less flattering as the one used by Saltzman, but it conforms better to the literature). 
Note that ice volume and ocean temperature are anomalies to the long-term (tectonic) average, and $\mu$ is calculated assuming $\mu_0=253~\textrm{ppmv}$. A similar figure was shown in the original article by Saltzman and Maasch \cite{Saltzman91sm}.}
%\end{minipage}
\label{fig:Saltzman15}
\end{figure}

Figure \ref{fig:Saltzman15} reproduces the original solution \cite{Saltzman91sm}, using the parameters published at the time. As in the original publication, the solution is compared with  Imbrie's $\delta^{18}$O-stack \cite{imbrie84} interpreted as a proxy for ice volume, but we have added here the CO$_2$ record extracted from the Vostok and EPICA (Antarctica) ice cores \cite{Luethi08aa,petit99}. 
Clearly the agreement between the model and the observations is seductive, but perhaps we should not be surprised by this, given that, many poorer models might have been rejected, and that the parameters in this model have been tuned in some manner to optimise the fit. 
Several authors have also argued that it is difficult to chose among models with similar dynamical behaviours but built on different interpretations of the climate system's functioning \cite{Roe99aa,tziperman06pacing}. So are such models useful ?

Let us first observe that while this question is a pressing one for phenomenological  models, it also concerns large general circulation models models (GCMs). Once assembled, GCMs are `tuned' to capture major characteristics of climate such as the overturning cell or the global mean temperature (e.g. \cite{jones05famoustuning}). This tuning is an effective way of incorporating macroscopic information in the model, but then this information  can no longer be treated as an emergent feature, i.e.\ an opportunity for validation, and a reason for confidence in prediction.  In philosophy, this conundrum is sometimes termed the problem of \emph{prediction versus accommodation} \cite{Hitchcock04aa,Lipton05aa}.

We believe that the difficulties of validation and prediction can be mitigated by the use of a formal framework in which all choices are clearly specified, and which proceeds according to well-established rules.  We will use a Bayesian statistical approach for this purpose.  This approach has its roots in the  works of Bayes, Laplace and Bernouilli, who were looking for ways to augment their knowledge of uncertain quantities such as initial conditions or parameters, by means of observations \cite{Jaynes79aa}. Rougier (2007) \cite{Rougier2007Probabilistic-i} explains how Bayesian inference methods may be applied to the problem of climate prediction.  We reprise this explanation here, in a slightly more general form, and adapted  to the problems posed by palaeoclimate time-series analysis. In particular, we  explicitly consider the dynamical aspects of climate,  the evolution of which is described by time-differential equations.

Before embarking on the mathematical details, it is useful to recall two aspects inherent in complex system modelling, as introduced in section \ref{remarks}. First,  by focussing on certain modes of climate variability we ignore a large body of information, such as its synoptic variability, or, e.g.: the occurrence of a particular volcanic eruption at a particular time. This  causes prediction errors that  have to be incorporated stochastically, since they are uncertain (we will here neglect the epistemological distinction between stochastic forcing and error).
Thus \textit{validation} consists in verifying that the complete set of model assumptions is compatible with observations, that is to say both the structure of physical model, \emph{and} the way in which we incorporate its limitations. In general it is not possible to test either of these two parts in isolation.
 Second, our physical model is constructed using parameters that cannot in practice be deduced from our knowledge of microscopic interactions. In our case,  these occur in  conjectures about the mathematical expressions of carbon, ocean and ice-sheet feedbacks, which are to be \textit{calibrated}, or tuned, by reference to observations.  Thus we must treat these parameters as uncertain, in order that we can learn about them.

 Prior to the present study Hargreaves and Annan \cite{hargreaves02} presented a Bayesian calibration of a slightly different version of the Saltzman and Maasch model. 
Our treatment addresses an acknowledged limitation of their approach,
which is that they treat their model as perfect, except for uncertainty
about the model-parameters.  This is seen in their likelihood function,
in which they compare the model trajectory at a given choice of
parameters with the observed trajectory, and scale by observational
error (which is assumed to be independent across observations).  We
explicitly include a term in the model to account for the limitations of
the model in describing the climate state at time $t+dt$ given the
climate state at time $t$.  Unlike Hargreaves and Annan, we will not be able to give a
simple closed-form expression for the likelihood function of the model
parameters, because once the discrepancy is included in this way, the
joint distribution of the parameters and the evolving state vector has a
very complicated structure, reflecting the underlying model.

 Now we outline the mathematical treatment.  Denote by $X(t)$ the state of the climate at time $t$, and by $X = \Set{ X(t) }$ the climate process over the period of interest.  We do not observe $X$ directly, but some proxy for it, which we denote as $Z$, and we denote the value of the observations as $z$.  Typically the observations in $Z$ are irregularly-spaced in time, and the relationship between $X$ and $Z$ is many-to-one; for example, $Z$ might respond to a subset of the components of the state vector, or to a subset of spatial locations.

 Our inference about $X$ based on the observations $Z = z$ can be expressed in probabilistic terms as $\pi ( X \given Z = z)$ where $\pi(\cdot)$ denotes a density function identified by its arguments, and `$\vert$' denotes `conditional upon'.  In other words, we would like to infer the density function of $X$ conditional upon the fact that $Z = z$.  Using the rules of the probability calculus, we can represent this inference in a way that is more amenable to calculations,
\begin{equation}
  \label{eq:PrXP}
   \pi(X \given Z = z) \propto \pi( Z = z \given X)\, \pi(X) ,
\end{equation}
by applying Bayes's Theorem---the missing constant is the reciprocal of $\pi(Z = z)$.  The density $\pi(Z = z \given X)$ describes the forward relationship between the climate state vector and the measured proxy: we will come back to this in section~\ref{Bayesian}.

The density $\pi(X)$ in \eqref{eq:PrXP} describes our uncertainty about $X$, and we introduce a physical model to help assess this density.  In our case, we have a deterministic model for the evolution of $X(t)$, which is specified conditional upon knowledge of the parameters $\psi$.  This model is written
\begin{equation}
  \label{eq:2}
  \frac{\rd X(t)}{\rd t} = f\big( X(t), t, \psi \big) .
\end{equation}
However, we are aware that this model has many limitations: some we know about, and \emph{could} include but choose not to, others we do not even know.  To capture the gross effect of these limitations, we introduce a stochastic component, writing our model in the It\^{o} form
\begin{equation}
  \label{eq:Ito}
  \rd X(t) = f\big( X(t), t, \psi \big) \, \rd t + \Sigma^{1/2}\, \rd W(t)
\end{equation}
where $\Sigma$ is a variance matrix that we must specify, and $W(t)$ is a vector of independent Brownian motions.  $X \given \psi$ is now a stochastic process, and consequently $X$ is a more complicated stochastic process, allowing $\psi$ to be uncertain.  However, both sources of uncertainty are necessary, because even if we knew $\psi$, we would not expect our model to do a perfect job of replicating the true system values $X$; and in fact we do not know $\psi$.  Therefore to account for the limitations of our model we will need to specify a density function $\pi(\psi)$ \emph{and} a variance matrix $\Sigma$.  While neither of these tasks is easy, it is better to do the best we can, than to accept obviously wrong choices, such as asserting that $\psi$ is known, i.e.\ $\pi(\psi) = \delta(\psi - \psi_0)$ for some $\psi_0$ that we specify, or that the model is perfect modulo our uncertainty about $\psi$, i.e.\ setting $\Sigma = \bs{0}$.

Once we introduce our physical model we introduce the new set of uncertain quantities, $\psi$.  Hence \eqref{eq:PrXP} becomes
\begin{equation}
  \label{eq:XFth}
  \begin{split}
    \pi(X, \psi \given Z = z)
    & \propto \pi( Z = z \given X)\, \pi(X, \psi) \\
    & = \pi( Z = z \given X)\, \pi(X \given \psi) \, \pi(\psi) ,
  \end{split}
\end{equation}
where in the first line we assert that $X$ is \emph{sufficient} for
$Z$, i.e.\ $\psi$ is not relevant to $Z$ when we have $X$, and the
rearrangement in the second line is a standard probabilistic
relationship.  The second term in the second line is the density
function of the stochastic model, \eqref{eq:Ito}, while the third term
is our prior assessment of uncertainty about $\psi$.  In practice,
we will not be able to evaluate the density function of
\eqref{eq:Ito}, but we will be able to sample from it, which is
sufficient for a numerical approximation of \eqref{eq:XFth}.

\begin{figure}
\begin{displaymath}
  \xymatrix{
    & z_{t_p-2} & z_{t_p-1} & z_{t_p} & & \\
    {\hphantom{X(t_p-3)}} \ar[r] & X(t_p-2) \ar[u] \ar[r] & X(t_p-1) \ar[u] \ar[r] & X(t_p) \ar[u] \ar[r] & X(t_p+1) \\
    & & \psi \ar[ul] \ar[u] \ar[ur] \ar[urr] & &
  }
\end{displaymath}
\bigskip
\label{fig:DAG}
\caption{Directed Acyclic Graph of the inference, where $t_p$ is the
  last time-point for which we have proxy measurements.  The process
  $X(t) \mathrel{\vert} \psi$ is Markov, allowing us to process the
  measurements sequentially in time.}
\end{figure}
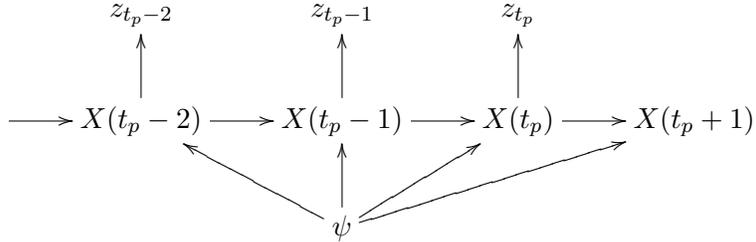

Eq.~\eqref{eq:XFth} is completely general: it requires no special conditions on the physical model, or on our probability assessments.  However, our dynamical stochastic model \emph{does} have special properties that we can exploit, notably that $X \given \psi$ is a Markov process, so that
\begin{equation}
  \pi \big( X(t+1) \given X(t), X(t-1), \dots, \psi \big) =   \pi \big( X(t+1) \given X(t), \psi \big) ;
\end{equation}
this can be summarised as stating that the past is irrelevant to the future, given the present.  Miller \cite{citeulike:1604706} describes it thus: ``Markov processes: the ordinary differential equations of the stochastic process realm'' (p.~18).  This dramatically simplifies the task of sampling from $\pi \big( X, \psi \given Z = z \big)$, because it allows us to proceed sequentially through time.  The Markov structure can be visualised using a Directed Acyclic Graph (DAG), as shown in Figure~\ref{fig:DAG}.  In this DAG, the middle row is expressed by \eqref{eq:Ito}, plus, if required, the initial density function $\pi\big( X(0) \big)$: we will not require $\pi\big( X(0) \big)$ because we will spin-up the dynamical model from its long-term (hereafter termed `tectonic') mean.  The bottom row is expressed by $\pi(\psi)$.  The connections between the middle and top rows express the forward relationship from the climate state vector at time $t$ to the proxy measurement at time $t$, and we have treated the climate state at time $t$ as sufficient for the proxy measurement at time~$t$.

Finally, we turn to the difficult issue of choosing $\pi(\psi)$ and $\Sigma$.  As we have already stressed, the challenge here is not to specify the `right' choices, but, at least initially, to avoid poor choices.  We regard both $\pi(\psi) = \delta(\psi - \psi_0)$ and $\Sigma = \bs{0}$ as poor choices.  It is very hard to make `not poor' choices directly, because neither $\psi$ not $\Sigma$ have clear operational definitions.  We reject out-of-hand the notion that any choices can represent `pure ignorance'.  A better approach is to make choices on the basis of their implications for the process $X$.  In other words, for each candidate choice we simulate trajectories of $X$ and consider whether, in our opinion, they provide a reasonable representation of our uncertainty about the evolution of the climate state.

\section{Bayesian calibration of SM91 and prediction using the particle filter\label{Bayesian}}

We now come back to system (\ref{SM91}) and operationalize the considerations of section \ref{bayes} to determine model parameters and provide predictions. Let us observe that certain parameters of SM91 may be estimated by means of  large numerical models of the atmosphere, ocean and ice sheets. 
We conserve the values proposed by \cite{Saltzman91sm} and ignore the time being the uncertainty on these parameters.
The $a_1$, $b_i$ and $c_1$  are far more difficult to estimate with general circulation model experiments and the sole prior constrain we have on these parameters is that they are positive. However, most instantiations of these parameters may be rejected by reference to the spectrum estimates of palaeoclimate data. Indeed, one may observe that (\ref{SM91}) has three fixed-points. Keeping in mind that the longest dominant period of the forcing is 41,000 years, the model has no chance of generating 100,000-year long cycles as those seen in the data, as long as the model response is quasi-linear around a fixed-point. We therefore need a mechanism to eject phase-space trajectories out of the fixed points in order to  get them to explore the entire phase space. One solution is simply to require all three fixed points to be unstable for zero forcing \cite{MAASCH90aa}.

The condition is  met when the $a_i$, $b_i$ and $c_i$ are set to the values published in \cite{Saltzman91sm} plus or minus roughly 20 \%. We therefore use this information to build prior distributions for these parameters (cf. Table~1) (a more formal construction of the priors is planned as a next step of the present project).

\begin{figure}
%  \centering
  \begin{minipage}{0.55\columnwidth}
  \includegraphics{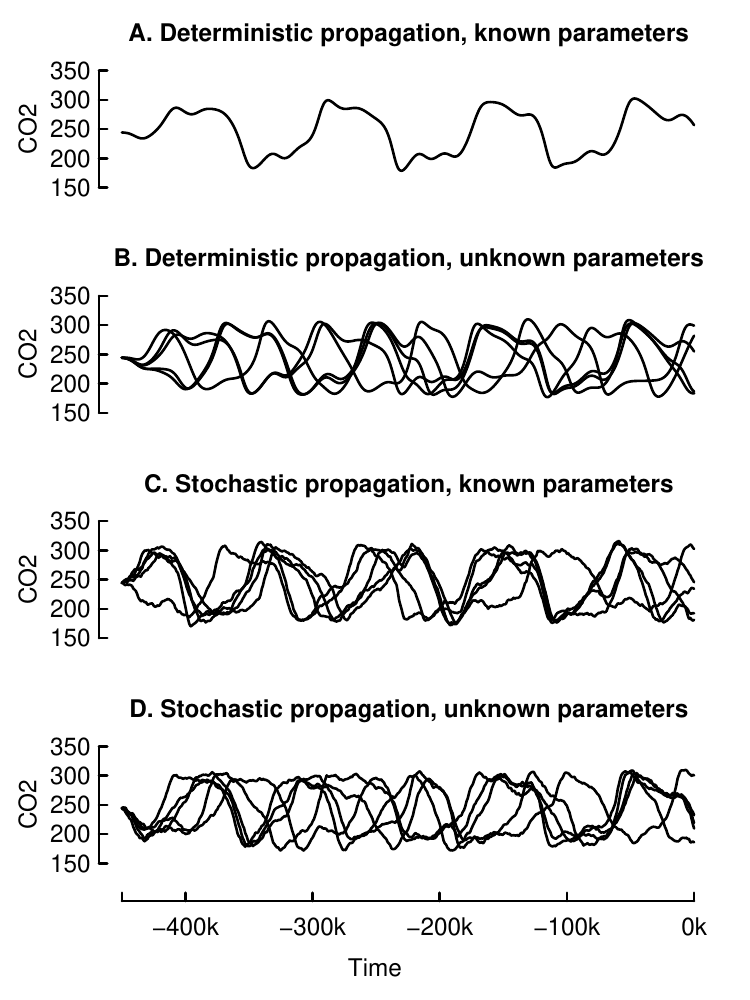}  
  \end{minipage}\hfill
  \begin{minipage}{0.40\columnwidth}
  \caption{Sample paths for \COtwo\ , based on choices for $\pi(\psi)$
    and $\Sigma$.  The state vector was started at $-450$~kyr, using the same initial value. Panel~A: deterministic propagation, and original choice of parameters $\psi$.  Panel~B: deterministic propagation, but for a random  sample of five different choices for $\psi$.  Panel~C: stochastic propagation (five samples), and original choice of $\psi$.  Panel~D: Same sample of $\psi$ as Panel~B, but with stochastic propagation.}
  \label{fig:spaghetti}
  \end{minipage}
\end{figure}
\begin{figure}[t]
\includegraphics{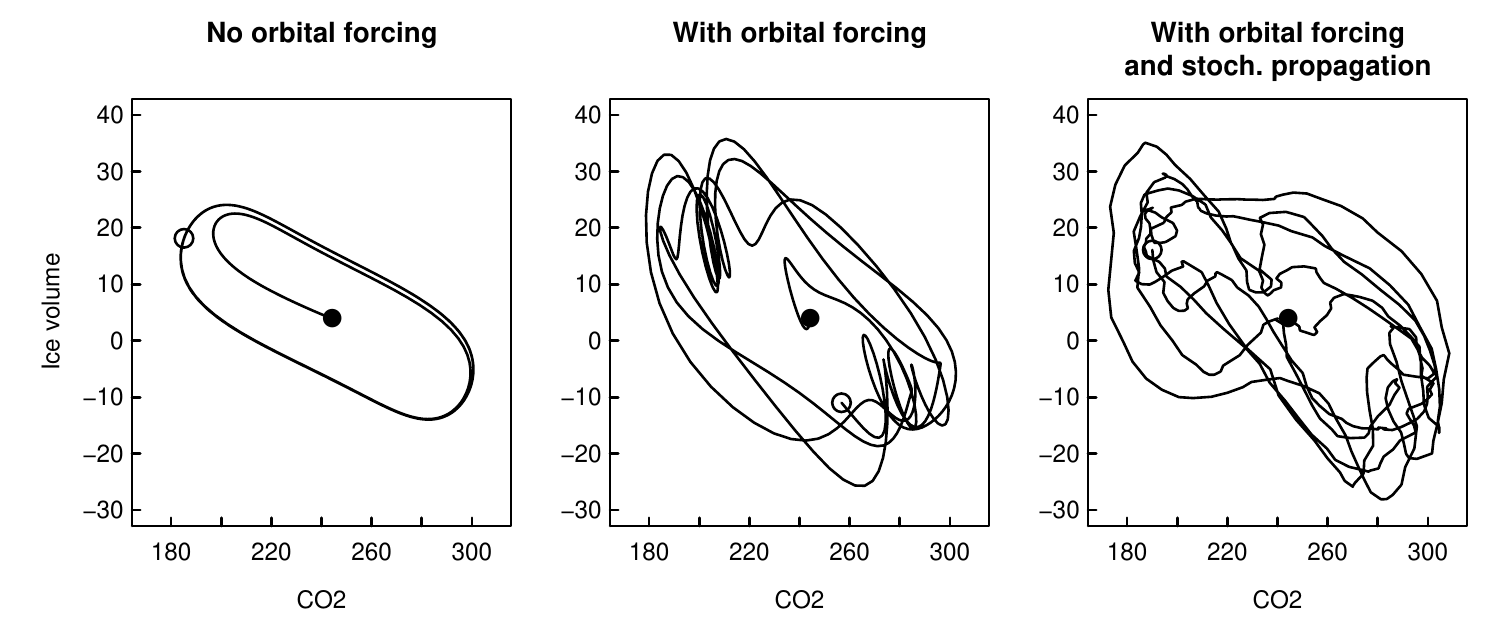} 
\caption{Phase-space diagrams of trajectories simulated with the SM91 model from $-450$~kyr to 0~kyr, using the standard choice of parameters $\psi$. The start and end of the trajectories are denoted by the black and white circles, respectively. The model exhibits a limit cycle in absence of external forcing, with a trajectory that resembles that obtained with data (Figure \ref{fig:phase}). The astronomical forcing adds a number of degrees of freedom that complicates the appearance of the phase diagram. The trajectory now crosses a number of local instability points around which a small stochastic forcing may have large effects.  }
\label{fig:traj}
\end{figure}

Figure~\ref{fig:spaghetti}A replicates the astronomically forced-evolution of \COtwo\ starting from $-450$~kyr using the tectonic mean as initial conditions. Panel~B shows five model realisations, starting from the same initial conditions but with different parameters sampled within the prior distribution. All paths show a typical oscillation of the order of 100-kyr---characterised by a hysteresis in the Ice vs \COtwo\ space (Figure~\ref{fig:traj})---
but they quickly split into groups that are out of phase with each other. The reason for this behaviour is to be found in the ability of the system to synchronise on the astronomical forcing, like heart beats  with the pulses of a pace-maker. The synchronisation process is here not trivial because the astronomical forcing is aperiodic (compare the autonomous and forced phase-space trajectories  on Figure~\ref{fig:traj}). In particular, small changes in model parameters may cause synchronisation slips near locally unstable points. This explains why originally close trajectories may become out of phase.

An extra layer of complication is added by the stochastic perturbation of the deterministic model. Consistently with the formalism developed in section \ref{bayes} we consider here additive perturbations of variance $\Sigma$. 
%The adopted value for $\Sigma$ (Table 1) may be justified in two ways. It may first be observed that with this value the Fourier-space behaviour of the model is consistent with palaeoclimate data estimates. Second, as shall be shortly seen, the chosen value of $\Sigma$ also provides posterior \COtwo\ and Ice volume estimates that are consistent with observations.\footnote{These two justifications refer to two different interpretations of the stochastic propagation term. The first justification considers that the stochastic term parameterises processes that we know occur---like weather fluctuations---but cannot be modelled otherwise than through their statistical distribution. This is the approach followed by Hasselmann \cite{hasselmann76} and pursued by Saltzman \cite{saltzman90sm} in the palaeoclimate context; the second justification is rather about the fact that we adequately model our \textit{uncertainty} on the system dynamics.}

\begin{table}[*p]
  \centering
  \framebox{
    \begin{minipage}{1.0\columnwidth}

  \caption{Parameters of the physical model, choices for the stochastic
    components, and the simulation method.  }
  \begin{enumerate}\addtolength{\itemsep}{2ex}
    
  \item Fixed (non-stochastic) model parameters:
    \begin{center}
    \begin{tabular}{lcc}
      $k_\mu$             & $11 / 253$ & \degC/$\textrm{ppmv}$\\
      $k_R$               & $0.08$     & $10^{18} \degC/(\textrm{Wm}^{-2})$ \\
      $k_\theta$          & $0.5$      & $\degC / \degC$ \\
      $K_I$               & $10^{-4}$  & $1 / \textrm{yr}$ \\      
      $K_\theta$          & $2.5 \times 10^{-4}$ 
                                       & $1 / \textrm{yr}$ \\      
    \end{tabular}
    \end{center}

    \item Stochastic model parameters ($\psi$), central values:

    \begin{center}
    \begin{tabular}{lcc}
      $a_1$ & $8.7 \times 10^{-4}$      & $10^{18}\text{kg}/\text{\degC\ yr}$ \\
      $b_1$ & $1.3 \times 10^{-4}$      & $1 /\text{yr}$ \\ 
      $b_2$ & $1.1 \times 10^{-6}$      & $1 /(\text{ppm yr})$ \\
      $b_3$ & $3.6 \times 10^{-8}$      & $1 / (\text{ppm}^2 \text{ yr})$ \\
      $b_\theta$ & $5.6 \times 10^{-3}$ & $\text{ppm} / (\text{\degC\ yr})$ \\
      $c_1$ & $1.2 \times 10^{-5}$      & $\degC / (10^{18} \text{kg} \text{ yr})$ \\
    \end{tabular}
    \end{center}
    \medskip

    For $\pi(\psi)$, the components of $\psi$ are treated as
    independent, and each component assigned a log-normal distribution
    with mean equal to the central value and a coefficient of
    variation of $20\%$ (i.e.\ the standard deviation is two tenth of
    the mean).

  \item Stochastic propagation variance ($\Sigma$) for the state
    vector $X = (I\, \text{$10^{18}$kg}, \mu\, \text{ppm}, \theta\,
    \degC)$:
    \begin{displaymath}
      \Sigma = 100^{-2}
      \begin{pmatrix}
        0.5^2 & 0 & 0 \\
        0 & 5^2 & 0 \\
        0 & 0 & 0.5^2
      \end{pmatrix} .
    \end{displaymath}

  \item Simulation method.  The stochastic differential equation was
    simulated using the stochastic Euler method, using a time-step of
    $100$~years.

  \end{enumerate}

    \end{minipage}
    }
\label{tab:choices}
\end{table}
Figure~\ref{fig:spaghetti}C shows five sample model trajectories with stochastic forcing, but with the original choice of model parameters. Again, trajectories split along different out-of-phase groups, but compared to Figure~\ref{fig:spaghetti}B they are also less smooth. The combination of both parameter uncertainty and stochastic propagation is shown on panel D. 

The above clearly shows that sequential state and parameter data-estimation techniques will be needed to provide predictions in the time domain, like, for example, the timing of the next glacial inception. The complications introduced by the highly non-linear character of the model dynamics and our uncertainty on parameters are such that there currently is no practical algorithm to solve this problem (this is a very active area of research in Statistics). Here, we used the  approximation proposed by Liu and West \cite{lw01} based on a \emph{particle filter}. 
It works as follows:
At each time step, the algorithm considers $n$ (here : $n = 50000$) samples of model parameters and states called `particles'. Particles are propagated forward in time according to a three-step process: (1) forward integration of the stochastic equations; (2) particle weight estimation as the product of prior weight and a likelihood taking into account the observation and its uncertainty; and (3) particle resampling in order keep a set of particles that all have approximately the same weight (importance resampling). We emphasise that this algorithm is a filter and not a smoother. Consequently, posterior estimates at any time $t$ are only informed by data prior to time $t$. 

\begin{figure}
\includegraphics{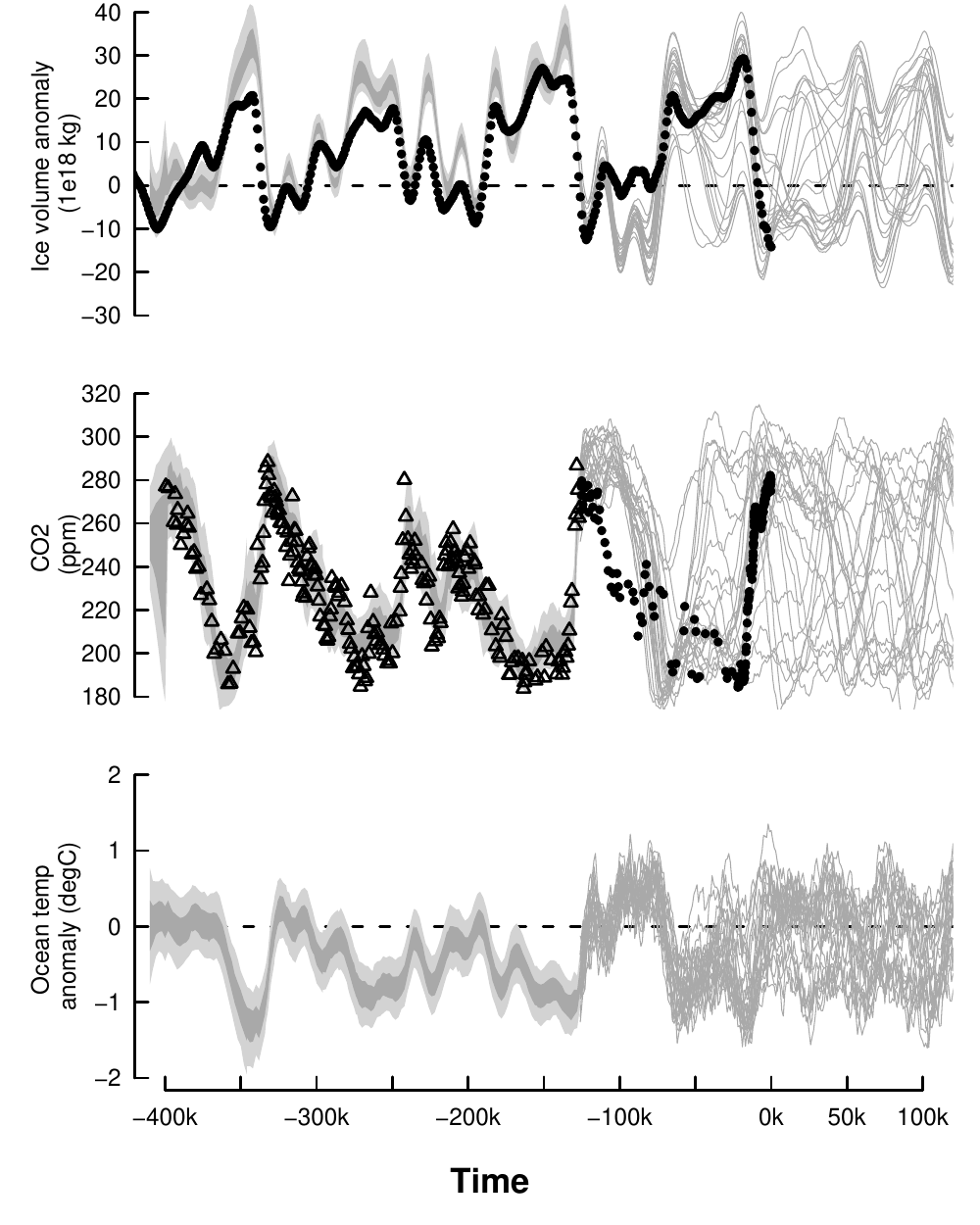}
\caption{
Stochastic propagation of the SM91 model (initialised at $-450,000$ years with priors as defined in the text), constrained by ice-core CO$_2$ data until $-126$~kyr \cite{petit99}. The training dataset is represented by triangles. Circles are validation data not used for model training. This includes (top) SPECMAP foraminifera data scaled as in Figure \ref{fig:Saltzman15} and \COtwo\ data after $-126$~kyr. 
The filtering distributions of the state through the training period are represented by shades, dark and light gray representing  the $[25^\textrm{th};75^\textrm{th}]$ and $[5^\textrm{th};95^\textrm{th}]$ quantiles of the particle weighted distributions, respectively. Thin lines after $-126$~kyr years represent $25$ simulated predictive trajectories, started at $-126$~kyr using initial conditions and parameters sampled from the posterior estimates at that time. }
\label{fig:filter126}
\end{figure}

Figure~\ref{fig:filter126} shows one realisation of the filter. The model is initialised at $-450$ kyr. The proxy ($Z$) is the \COtwo\ measured in ice-core air bubbles at Vostok, Antarctica \cite{petit99}, but for updating we only use data  between $-400$ and $-126$ kyr. 
CO$_2$ presents the advantage of being included in the climate state vector. We simply need to specify an observation error, here assumed to be Gaussian with 20~ppmv standard deviation. Furthermore, the GT4 time scale covering the interval $-400\,$kyr to present is only weakly constrained by orbital assumptions, with only two tight-points ($-110\,$kyr and $-390\,$kyr).

This first experiment allows us to validate both the model and the particle filter. 
It constitutes an interesting test to the model because $-126$~kyr is the end of the penultimate deglaciation, which puts us in conditions analogous to the early Holocene. 
First, the model correctly captures ice volume variations, here indicated by the SPECMAP stack record \cite{imbrie84} (see the Figure caption for details).
The way the model tracks ice volume data is surprisingly good. It is not perfect, but the SPECMAP stack is not a perfect proxy for ice volume either because it is partly affected by ocean temperature. For example, the ice volume maximum at $-230$~kyr (this is event 7.4 in the SPECMAP nomenclature) is overestimated by the model, but there is also evidence that the event is underestimated by SPECMAP \cite{lisiecki05lr04,Siddall03aa}. Second, model predictions for climate evolution after $-126$~kyr are overall consistent with data. These predictions are shown by the thin gray lines. They are obtained as model realisations sampled from the latest state and parameter estimates at $-126$~kyr.
In particular the model succeeds in predicting the draw-down in \COtwo~and the large ice mass increase around $-70$~kyr. Model predictions then split but 60 \% are consistent with ice volume above $20~10^{18}$ kg of ice at the last glacial maximum, and 36 \% are consistent with the more stringent prediction of ice volume above $30~10^{18}$ kg of ice at that time.
By contrast, the late decrease in \COtwo\  around $-90$~kyr as compared to the data constitutes a severe weakness. The relatively weak predictability of \COtwo\ is also noteworthy. Poor predictability is the price we pay for correctly accounting for both model and data uncertainties. 

To the extent that Figure \ref{fig:filter126} seems to us to 
be a reasonable description of our uncertainty about the
dynamics of Quaternary climates, we regard the combination of our physical model
and our statistical choices as `valid'. We are now in a position to provide a prediction for climate evolution after the early Holocene. Conservatively, any data after $-8$~kyr years is ignored since they may have been contaminated by human influence according to the early anthropogenic hypothesis. The forecast therefore starts at $-8$~kyr.

\begin{figure}
\includegraphics{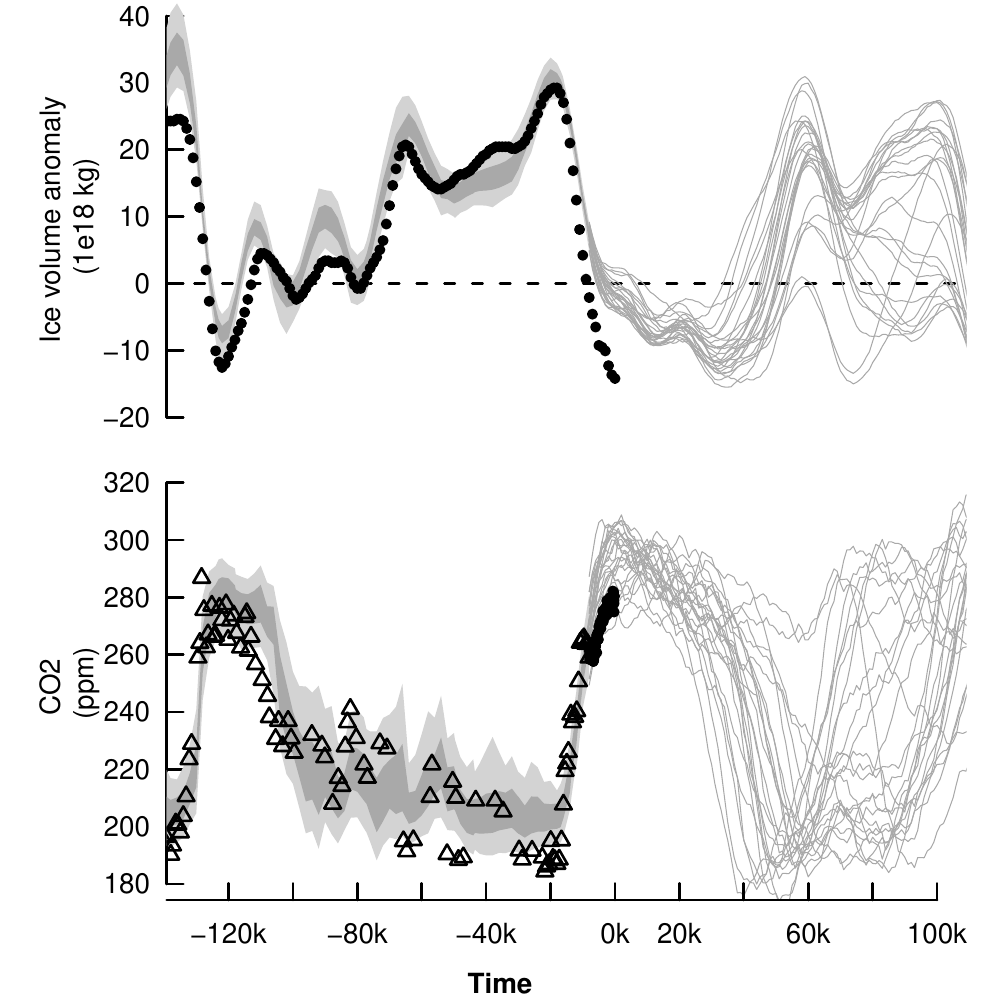}
\caption{
Same as figure \ref{fig:filter126} but with training until $-8$~kyr. Data after $-8$~kyr are conservatively neglected here as a test of Ruddiman's hypothesis, according to which humans have been perturbing climate since that time. Predicted trajectories remain, though, consistent with a long interglacial with no glacial inception before about $40$ to $50$~kyr from now, and a maximum in ice volume at + $60$~kyr. }
\label{fig:filter8}
\end{figure}
Results are shown on Figure \ref{fig:filter8}. Again, the model provides a particularly good tracking of the ice volume record up to $-8$~kyr even though this record has not been used to train the model. The model prediction after the early Holocene is then a gradual melt-down of ice, further increase in CO$_2$ and then a slow decrease up to $+40$~kyr, at which time glacial inception is expected with good confidence.  Predictability after that point is poor. This prediction is consistent with the Berger and Loutre one \cite{Berger2002An-exceptionall}.  

At this stage of our research project, we still need to emphasise that the \COtwo\ dynamics during interglacial  periods are not satisfactorily reproduced by our model, as we have seen on Figure \ref{fig:filter126}. This  implies that we cannot yet be confident that this prediction of the glacial inception is well-calibrated, but given the model success in predicting the timing of peak glacial conditions, our prediction for ice volume maximum at +60~kyr is, we think, trustworthy (ignoring the possible effects of an anthropogenic perturbation).
As a straightforward extension of the present work it will also be informative to examine the period $-800\,$kyr to $-400\,$kyr, during which CO$_2$ variations exhibited a smaller amplitude. It is expected that the calibration parameters needed to accommodate this earlier part of the record will be different, and it would be insightful to interpret these differences.
 
As pointed out by one referee (P. Huybers), an issue may arise if structurally different models equally pass the validation test but yield incompatible predictions. In this case one must hope that an extra validation criteria will lead us to prefer one model to the other.  The contrary will be an indication that model discrepancy was underestimated. 

\section{Conclusion} 
Behind this paper is the message that climate modelling is not and should not be a merely technological question, to be solved with larger and more complicated climate models, and faster computers. Of course, such models skillfully predict many complicated aspects of atmosphere and ocean dynamics; in that sense they are important and useful. Yet they represent just the far end of a spectrum of models, all of which have a role to play.

We have emphasised the use of phenomenological models to capture the wealth of the data in the palaeoclimate record.  These models are tractable, and a rich source of insights about climate behaviour on large spatial and temporal scales.  In their complexity, they are also beautiful mathematical objects.  But they present some serious challenges if they are to be used quantitatively, since,  with their reduced physics and their under-determined parameters, they make far greater demands on our inferential framework.  We have used Bayesian statistics to combine information from the observational record and the model behaviour, accounting for both parametric uncertainty and model limitations.

More generally, we have stressed again that the analysis of palaeoclimates is a truly multi-disciplinary experience, starting with the field scientists who retrieve and interpret the observational data, and then encompassing a wide range of applied mathematicians, including climate modellers, complex systems experts, and statisticians.  As we hope we have shown, palaeoclimates pose challenges in all of these areas, and questions as deceptively simple as `When do we expect the next glacial inception?' require us to operate at the frontier of our knowledge and technique.

\section*{Acknowledgements}
MC thanks the organisers of the 1st International Workshop on Data analysis and modelling in Earth Science held in Potsdam for their invitation and travel support.  Correspondence with Julia Hargreaves and James Annan is in part at the origin of this project. 
The comments of the editor Reik Donner and two reviewers, among whom Peter Huybers, are gratefully acknowledged. MC is supported by the Belgian National Fund of Scientific Research, JC is supported by the NERC-QUEST palaeoQUMP project. The collaboration between the authors was made possible by a grant offered by the British Council and CGRI/FNRS. All calculations were performed in the statistical computing environment R \cite{R-Development-Core-Team04aa}.

\renewcommand{\baselinestretch}{0.9} % or 2, or whatever
\small
\bibliography{/Users/crucifix/Documents/BibDesk.bib}
\end{document}